\documentclass[review]{elsarticle}

\usepackage{lineno,hyperref}
\usepackage{lmodern}
\usepackage{graphics}
\usepackage{layout}
\usepackage{amsmath}
\usepackage{textcomp}
\usepackage{amssymb}
\usepackage{amsfonts}
\newtheorem{remark}{Remark}
\usepackage{array}
\usepackage{color}
\usepackage{pstricks}
\usepackage{pst-node}
\usepackage{graphicx}
\modulolinenumbers[5]

\journal{Physica A}

\begin{document}

\begin{frontmatter}

\title{Schr\"{o}dinger type equation for subjective identification of supply and demand}
\author[mymainaddress]{Marcin Makowski}
\ead{makowski.m@gmail.com}
 \author[mymainaddress]{Edward W. Piotrowski}
\ead{qmgames@gmail.com}
\author[mysecondaryaddress]{Jan S\l{}adkowski}
\ead{jan.sladkowski@us.edu.pl}


\address[mymainaddress]{Institute of Mathematics, University of Bia\l{}ystok \\ K. Cio\l{}kowskiego 1M, 15-707 Bia\l{}ystok, Poland}

\address[mysecondaryaddress]{Institute of Physics, The University of Silesia,\\
     75 Pu\l{}ku Piechoty, Pl-41-500 Chorz\'ow, Poland}

\begin{abstract}
The present authors have put forward
a quantum game theory based  model of market prices movements. By using Fisher information, we present a construction of an equation of Schr\"{o}dinger type for probability distributions for relationship between demand and supply. 
Various analogies between
quantum physics and market phenomena can be found. 
\end{abstract}

\begin{keyword}
Fisher information\sep supply and demand\sep negative probability\sep market
\end{keyword}

\end{frontmatter}


\section{Introduction}
\label{intro} A considerable part of economics is focused on various types of  demand and supply curves  that represent a concise account of market situation \footnote{M.~ Blaug
 quotes at least a hundred of such diagrams \cite{blaug}.}. They are used for forecasting in statistical analyzes of various types.  Paradoxically, the shape of such curves scarcely results from a  measurement-like process. We focus on the issue of determining the shape of demand and supply curves. In the Information Theory Model of Markets approach put forward in ref. \cite{geom} these curves are parameterized by  the logarithm of the price  of the good we are interested in (here denoted by x). From the seller's point of view, the value  x is being offered for a unit of the good she/he wants to sell -- and the agent accepts  or rejects it (or vice versa)\footnote{A word of warning: although these situations  are apparently symmetric, in some contexts, they might require slightly different quantitative analysis, c.f.  \cite{qmg}.}). The agent can try to approximate the supply/demand curves for the good in question by performing an  analysis of previous market activities. In such case  these curves could be interpreted as probability distribution functions (pdf) of probabilities of accepting bids put forward by market. Therefore supply/demand curves define two random variables $x$ -- the logarithm of buying price and $y$ -- the logarithm of selling price. Both characteristics are subjective and often kept private because:
\begin{enumerate}
\item they depend on knowledge and abilities of the agent\\ and
\item results from some observer's  (not necessarily the buyer or seller) interpolation that involves "experimental" errors and interpretations.
\end{enumerate}

Re: 1) An agent that enters the game adopts a specific  strategy  to accomplish her/his aims. The agent' individualities and market performance make individual contribution to the market as a whole and determine effects of agent's activities.

Re: 2) The supply/demand curves parameters  are determined from a finite set of
(often scarce) historical data  by estimation in a selected a priori class of statistical distributions, or
                  values of higher moments of statistical distributions following from averaging over the sample.
                                   
We will use the second method because it does not involve arbitrariness in  pre-selection of some class of pdfs. Shapes of the curves are selected by a criterion based on minimization of information measures. In our analysis, we use Fisher's information which is a function of the characteristic probability (the expectation value of the logarithmic
gradient of density) \cite{Frieb,flip}:

\begin{align}\label{fish}
	I_{F}=\int f(x)\bigg(\frac{d}{d x}\ln f(x)\bigg)^2  dx \,.
\end{align}
For the convenience of notation we put $f(x)=\psi^2(x)$. Assuming that the probability density is defined by a real valued function we receive the following formula
\begin{align}
	I_{F}=4\int[\psi^{\prime}(x)]^2  dx \,,
\end{align}
where $\psi^{\prime}(x)=\frac{d}{d x}\psi(x)$.\\
\noindent
We consciously consider a wave function with real values  only  as this is sufficient for our purposes (description of the market price movements).  Such an  assumption was also used  in the model of a one-dimensional linear oscillator \cite{landau} to which we refer in our work.
We allow complex values  of $\psi(x)$ function. This can be useful in analyzing different (not considered in this article) effects.

\section{Analogies between supply/demand curves and quantum states}
One of the most basic economic laws states that supply and demand are monotonic functions of prices \cite{blaug}. 
There is an interesting probabilistic interpretation of the supply and demand curves \cite{sigmafi}:
\begin{eqnarray}
CDF_s(x)=\int_{-\infty}^{x}f_1(p)dp\,\,\,\,\,\,\,\,\,\,\,\,\,\,\,(supply)\,,\nonumber\\
CDF_d(x)=\int_{x}^{\infty}f_2(p)dp \,\,\,\,\,\,\,\,\,\,\,\,\,\,\,(demand)\,,\nonumber
\end{eqnarray}
where $f_1$, $f_2$ are appropriate probability density function, in general case they are different due to various properties of the market (monopoly, specific
market regulations, taxes).
The value of the supply function $CDF_s(x)$ is given by the probability of the purchase of a unit at
the price $\leq e^x$ (and analogously in the case of demand, for more details see \cite{sigmafi,Mak}). 

The agent defines its position on the market by choosing a specific supply/demand curve (strategy).
Despite different objects and terminology, supply/demand curves and wave functions (strategies in the terminology of game theory \cite{qmg}) have a lot in common. Consider the following examples. A (subjective) supply curve characterizes whole market but the very agent whose strategy is used to fix the curve. Analogously, electron's wave function reflects interactions with other objects with exclusion of this electron. Quantum states are "unknown" unless measured and measurements influence them - often in a dramatic way. One usually supposes that markets are efficient, that is, among others, cannot be influenced by a single agent's strategy. Measurements of a market state by an agent's transactions reveal information on market supply but her/his activity should not (ex definitione) affect the market. Both, agent and, say, particles can aggregate  but these groups (coalitions) should not loose their specific properties (quantum in case of particles and merchant for traders). A lot of other arguments can be stated to encourage the application of what can be called {\it quantum decision theory} to describe some aspects of market phenomena \cite{segseg}-\cite{zasada}. The terminology may sound sometimes strange but if one remembers that there are no underlying real quantum processes but only the underlying logical structure is addressed and interpreted. Therefore, there should be no misunderstanding.
\section{The Cram\'er-Rao inequality and quantum uncertainty bounds}
The Cram\'er-Rao inequality (lower bound) says that the variance of an (unbiased) estimator is bounded below by the inverse of the Fisher information \cite{Frieb}. 
Existence of inequalities of this type can be proven in a quite general setting and one should not be surprised to find them in statistics and finance theory. Elsewhere  \cite{geom}, we have shown that the natural measure of transaction profit is the logarithmic rate of return $x$ i.e. the logarithm of transaction price. One of the most important factors for market  strategies is  the profit uncertainty $\Delta x$, that is risk measured by  the second moment of the stochastic variable $x$, cf.~\cite{zasada}.  By simple reformulation  -- that results directly from Cauchy-Schwarz inequality --  the Cram\'er-Rao lower bound can be rewritten  in the form of the  uncertainty relation \cite{Frieb} :
\begin{equation}
\Delta (\frac{\partial S}{\partial x})\, \Delta x \,\geq\,1\,. 
\label{bond}
\end{equation}

The term $\frac{\partial S}{\partial x}$  is usually referred to as the derivative of  surprisal function $S(x)=-\ln\,p(x)$. We call the uncertainty of this quantity
{\it the Fisher information $I_F$}. The surprisal function measures (counts) events of small probability, the so called {\it black swans} \cite{taleb} but its derivative signals their appearance or disappearance. In the context of finance theory, we speak of buy or sell signals. The dispersion of the derivative  is big in situations when theoretical expectation of black swans is unclear. In such cases market does not follow a definite trend, suffers from sudden  and substantial volte-faces. On such turbulent market only cold blood and experience give chance for survival. In that sense, the Fisher information measures informative aspects of statistical distribution functions. But what is the meaning of the bound $(\ref{bond})$ for continuous models of market  transactions?
Below  we would look for an answer to the following question: {\it What are the elementary consequences for demand/supply perception under conditions of minimal market (Fisher) information?}

\section{Market activities minimize information}

 The subjectiveness of the analyses  suggests that market dynamics should be interpreted as  market tendency  to minimize the  information revealed about itself. There are at least two sorts of arguments for minimizing information about markets. The first class of arguments (logical) is based on the famous Laplace principle of indifference: if we have no information on measures of probabilities of elementary events then we should  treat all of them on the same footing (ie as equivalent). The second one  follow from  the {\em no  free lunch} principle. One should not expect anything else - more information involves higher costs on the revealing information side ({\it information is physical} \cite{land}). Consider the following variational problem for market heuristic (subjective) curves of supply (demand)\footnote{See Extreme Physical Information  principle (EPI) \cite{Frieb}.}.
Let $f(x)$ be the PDF of random variables $x$ (the logarithm of buying price) with the mean value $m$  and the corresponding risk $r$:

\begin{equation}
\label{brzegi}
1\,=\,\int_{-\infty}^{\infty}{f(x)}dx\,, \,\,\,m=\int_{-\infty}^{\infty}{x\, f(x)}dx\,, \,\,\,r=\int_{-\infty}^{\infty}{(x-m)^2\,f(x)}dx\,.
\end{equation}

We are looking for a function $f_{\min}(x)=:\psi^2(x)$ for which the $I_F$ takes the minimum value under condition (\ref{brzegi}). This problem can be solved by using Lagrangian multipliers $a$, $b$, $c$ and searching for a minimum of the functional: 
\begin{align}\label{fun}
	\int_{-\infty}^{\infty} F(\psi(x),\psi^{\prime}(x),x)\,dx\,,
\end{align}
 where 
\begin{align}
	F(\psi(x),\psi^{\prime}(x),x)=4[\psi^{\prime}(x)]^2 -(a\,+\,b\,x\,+\,c\,x^2)\psi^2(x). \nonumber
\end{align}
For the functional (\ref{fun}) the extremal\footnote{Because $\frac{\partial^2 F}{\partial {\psi^{\prime}}^2} >0$  this is the minimum (see e.g. \cite{oscy}).} function satisfies the Euler-Lagrange Equation \cite{Arnold}:
\begin{align}\label{roww}
	\frac{d}{dx} \left(\frac{\partial F}{\partial \psi^{\prime}}\right) - \frac{\partial F}{\partial \psi} = 0\,.
\end{align}
In our case, the condition (\ref{roww}) takes the form:
\begin{align}
	8 \frac{d}{dx} \psi^{\prime}\,+\, 2\,(a+b\,x+c\,x^2)\psi\,=\,0\,.\nonumber
\end{align}
Hence 
\begin{align}
	 -\frac{d^2\psi}{dx^2} \,-\,\tfrac{1}{4} (a+b\,x+c\,x^2)\psi\,=\,0\,.\nonumber
\end{align}
Substituting

\begin{align}
 a=8\,\varepsilon\,\mu-4\, x_0^2\,\mu^2,\,\,b=8\,x_0\, \mu^2,\,\,c=-4\,\mu^2\,,\nonumber
\end{align}
we get the following equation: 
\begin{align}\label{sch}
	 -\frac{d^2\psi}{dx^2} \,+\,\mu^2(x-x_0)^2\psi\,=2\varepsilon\mu\,\psi\,.
\end{align}
Multiplying both sides of the above equation by $\tfrac{1}{2\,\mu}$  and replacing a simple derivative with a partial derivative ($\psi$ is a function of one variable $x$) equation (\ref{sch}) takes the form: 
 \begin{equation}
-\frac{1}{2\mu}\frac{\partial^2 \psi}{\partial x^2}+ \frac{\mu}{2}(x-x_0)^2\,\psi\,=\,\varepsilon \,\psi\,.\nonumber
\end{equation}
After translation $x\mapsto x+x_0-m$ we get:
\begin{equation}
\label{zyta}
-\frac{1}{2\mu}\frac{\partial^2 \psi}{\partial x^2}+ \frac{\mu}{2}(x-m)^2\,\psi\,=\,\varepsilon \,\psi\,,
\end{equation}
\noindent
Let us stress that in the above derivation we assume that $f_{\min}(x)=:\psi^2(x)$ ($\psi$ -- real valued wave function) and we use the definition of Fisher information. No other assumptions are necessary. The procedure of minimizing the Fisher  information fulfilling the  conditions (4) led us to Hamiltonian in the form that we can observe in the equation (\ref{zyta}).
\begin{remark}\label{ryzyko}
Note that Fourier transform (FT) of equation (\ref{zyta}) lead to equation of the same type. This is a consequence of the following properties of the Fourier transform:

\begin{align}
	-\frac{d^2\psi(x)}{dx^2}\xrightarrow{FT} y^2\,\psi(y) & \,,\,\,\,\,\,\,\,\,\,\,\, \,\,\,\,\,x^2\,\psi(x)\xrightarrow{FT} -\frac{d^2\psi(y)}{dy^2}\,,\nonumber
\end{align}
where $\psi(y):=FT\,\psi(x)$.
\end{remark}

\begin{remark}
The term $(x-m)^2\,=(x-\langle x \rangle)^2$ of the equation (\ref{zyta}) will be called risk operator (selling risk or supply risk) because its expectation value (see (\ref{brzegi})) 
\begin{align}
 \langle (x-\langle x \rangle)^2\rangle = \int_{-\infty}^{\infty}(x-m)^2\,\psi^2 (x)\, dx \nonumber
\end{align}
corresponds to the variance of a random variable $x$ (one of the measures of risk considered in financial mathematics). The operator $-\frac{\partial^2}{\partial x^2}$ is associated with $(x-\langle x \rangle)^2$ by Fourier transform (see Remark \ref{ryzyko}). 
\end{remark}
We see that in this way we derived an equation of  the  known {\it Schr\"odinger  type}. All solutions of Eq. $(\ref{zyta})$ form a discrete family of functions (see \cite{Frieb}):
$$
\psi_n(x)=\sqrt{\frac{\sqrt{\mu}}{2^n n! \sqrt{\pi}}}\,e^{-\tfrac{\mu\,(x-m)^2}{2}}\,H_n(\sqrt{\mu}(x-m)),
$$
where $\varepsilon=\varepsilon_n=n+\tfrac{1}{2}$, for $n=0,1,2,...$\,.
The quantum-like character of the presented theoretical model follows from the very natural requirement of minimization of information revealed by markets.

When market is at ''local minimum of Fisher information'' one can easily prove that:
\begin{itemize}
\item the total supply/demand risk is discrete ($=n+\tfrac{1}{2}$),
\item the square root of PDF function for demand is the Fourier transform of  the square root of PDF function for supply (see eg \cite{qmg})\,,
\item  $y$ (the logarithm of selling price) is dual to $x$ (the logarithm of buying price) and
\begin{center}
$\Delta x \,\Delta y \,\geq\, 1$,
\footnote{The relationship between $x$ and $y$ is the same
as for position and momentum variables in quantum mechanics. Inequality results from Heisenberg's uncertainty principle.}
\end{center}
\item the Fisher information at its local minimum $I_{\min}\sim\varepsilon_n$ is an invariant of Fourier
transformation (supply/demand shift invariance).
\end{itemize}
\begin{figure}
\includegraphics[width=100mm, height=50mm]{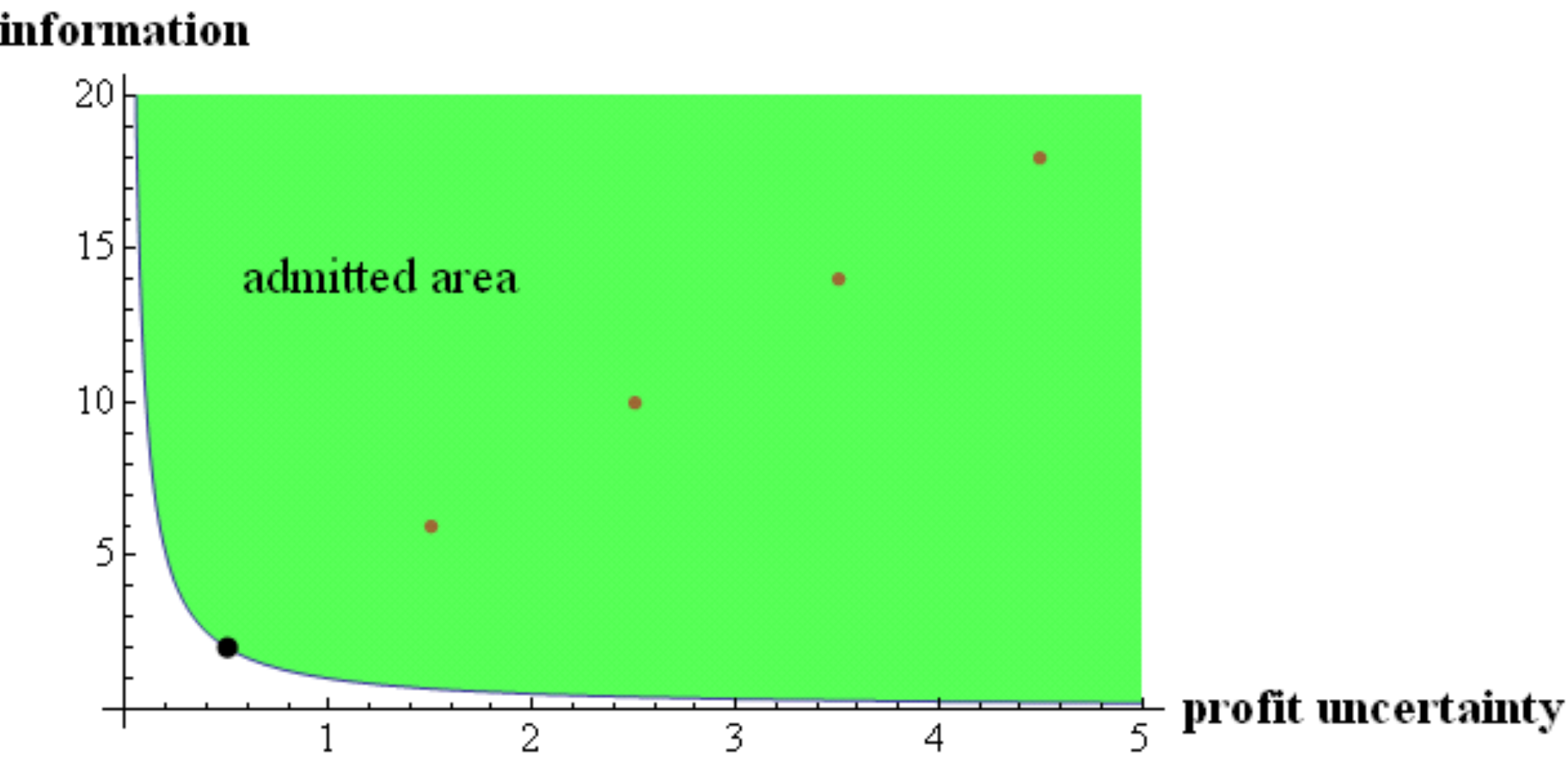}\phantom{.}
\caption{The local minima of Fisher information.}
\end{figure}
 

 The Fisher information takes the minimum value  $\varepsilon_n$ for the strategy $\psi_n$, that is  equal to the eigenvalue of  the operator defining the left hand side of Eq.$(\ref{zyta})$ that is the sum of the supply risk operator  $(x-\langle x \rangle)^2$ and $-\frac{\partial^2}{\partial x^2}$. It is tempting to call it {\it the minimum information operator}. Let us have a closer look at the second term. Any market adopting the minimal supply strategy $\psi_n(x)$ does not reveal additional information if it shows the demand  $\psi_n (y):=FT\,\psi_n (x)$\footnote{This is the assumption of our model, see \cite{PSS}. }, where $FT$ denotes the Fourier transform. This is because the quadratic form corresponding to  the minimum information operator is invariant with respect to the Fourier transform. The strategy that minimizes Fisher information on market is the same from the point of view of seller and buyer in the sense that they are mutually theirs Fourier transform. 
In the demand representation \cite{qmg}, $-\frac{\partial^2}{\partial x^2}$ takes the form $FT\, (-\frac{\partial^2}{\partial x^2}) \,FT=(y-\langle y \rangle)^2 $. This allows for its interpretation as the demand risk operator  and we have a connection between risk and information associated  with strategy: minimal information content of a market supply/demand strategy is equal to the sum of the related supply and demand risk (actually this is a sum of two noncommuting operators!). Therefore, it is natural to assume that (\ref{zyta}) is sort of a general
\textbf{risk balance equation}.

In general, the set of all market strategies contain a countable subset of local  minimum  strategies. Recalling the well known facts concerning quantum harmonic oscillator, we can claim that any strategy (actually its probability amplitude) can be approximated with arbitrary precision by a dense subset of amplitudes that minimize probability. Therefore, in principle, we can express any market strategy amplitude as a linear combination of amplitudes that minimize information about the market -- and the squared absolute values of the coefficients give probabilities that the results of the basic strategies would correspond to situations that the market adopts one of these locally minimal strategies.

We should remember that a linear combination of two strategies is  from the information point of view simpler than the corresponding mixed strategy (convex combination) and for mixed strategy we should also take the Boltzmann-Shannon entropy for the weights (coefficients) into consideration. Note that the invariance of our approach with respect to the supply-demand behavior is necessary because when you buy you exchange money for o goods and when you sell you actually exchange goods for money (and money is simple a preselected good).

\section{The Wigner functions of market}
The above results  can be presented in an elegant way with the help of the formalism of Wigner  functions defined on the common domain of variables $x$ and $y$ (the phase space):
$$
f_n(x,y)=\frac{1}{2\pi}\int_{-\infty}^\infty{\psi_n (x +\tfrac{s}{2})\,\psi_n(x-\tfrac{s}{2})\,\cos(s\,y)}ds\,.
$$
Conditional (fixed public price for buying or selling) demand and supply curves are depicted by the graphs of the following CDFs:
$$
CDF_d(\ln\,c)=\int_{-\infty}^{\ln\,c}{f_n(x=const.,y)}dy\,,
$$
$$
CDF_s(\ln\,c)=\int_{-\infty}^{\ln\,\tfrac{1}{c}}{f_n(x,y=const.)}dx\,,
$$where $c$ denotes  the price of the good in question.
Unfortunately,  only for $n\negthinspace=\negthinspace 0$ ($\varepsilon\negthinspace=\negthinspace\tfrac{1}{2}$) the Wigner function  is positive-definite (the surface of gaussian shape).
\begin{figure}
\begin{center}
\includegraphics[width=55mm]{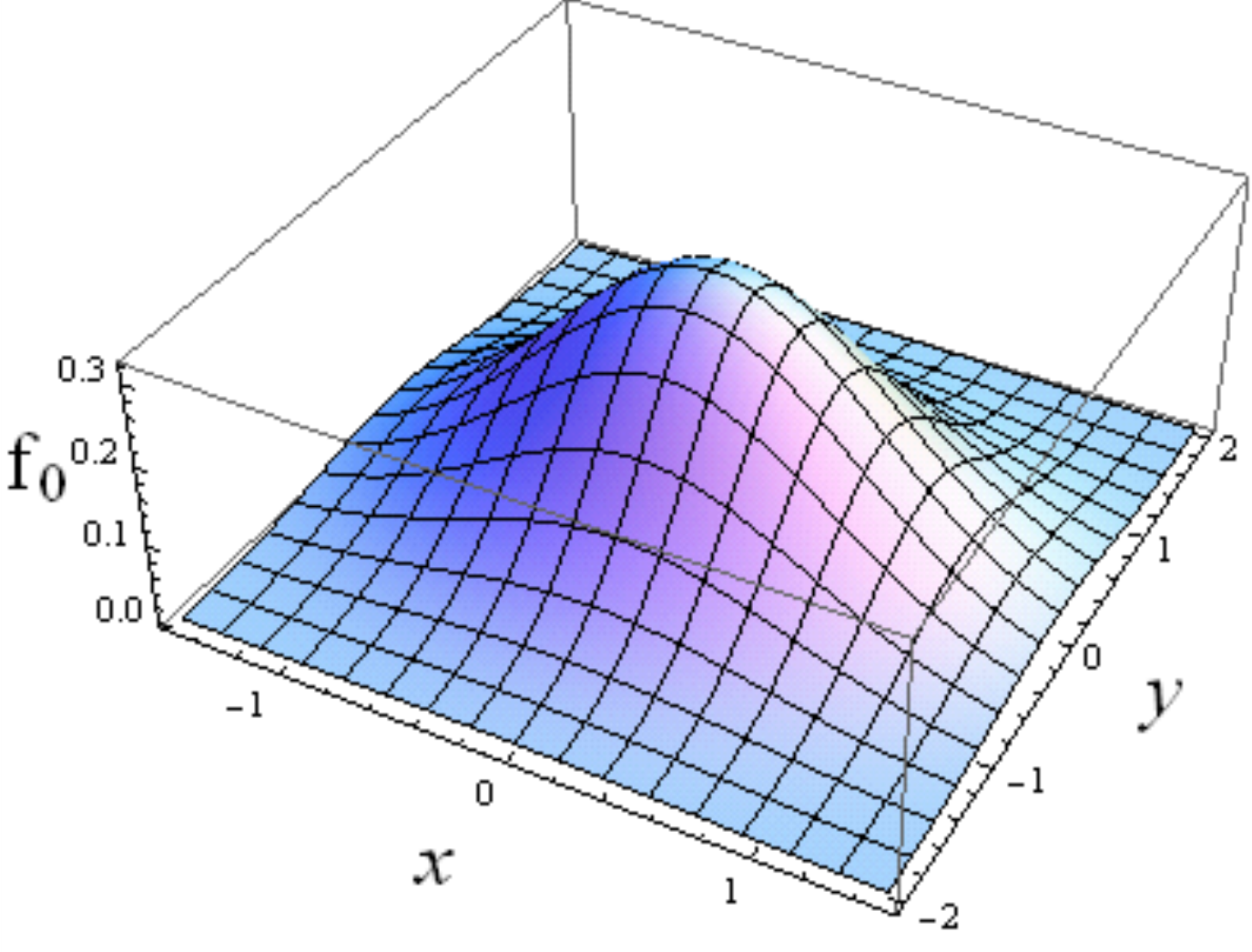}
\includegraphics[width=55mm]{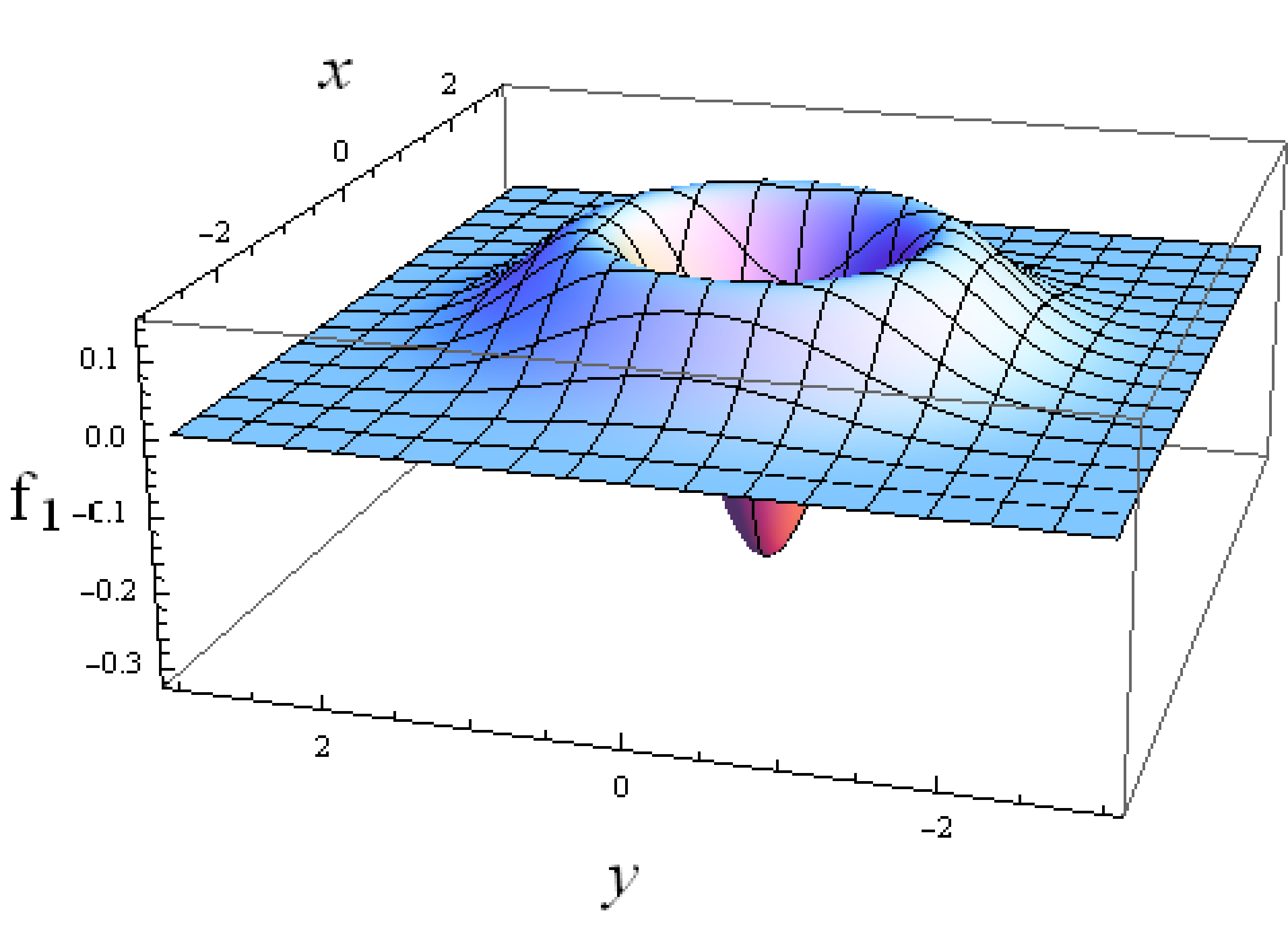}\\[3.5ex]
\includegraphics[width=55mm]{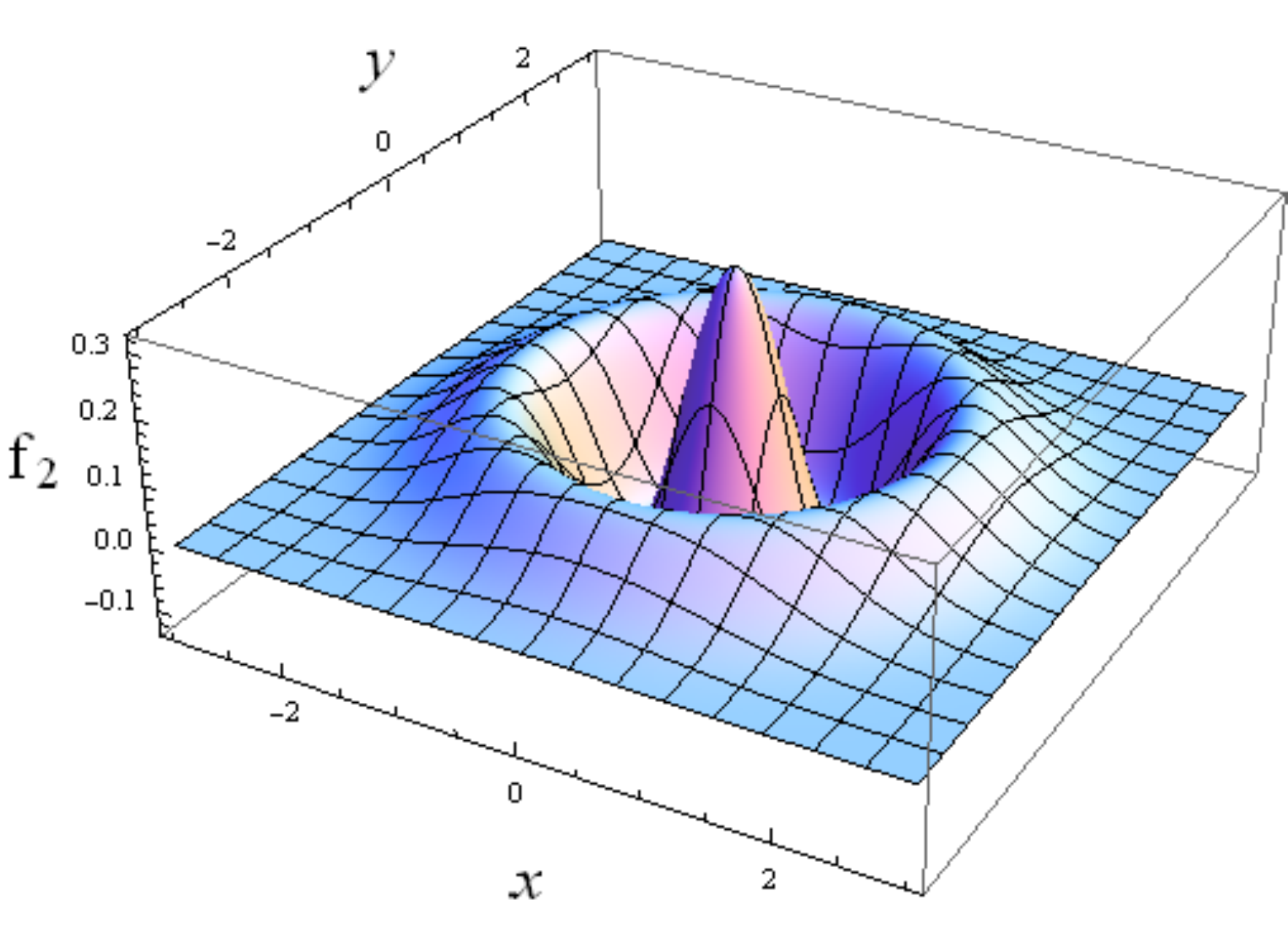}\quad\quad
\includegraphics[width=55mm]{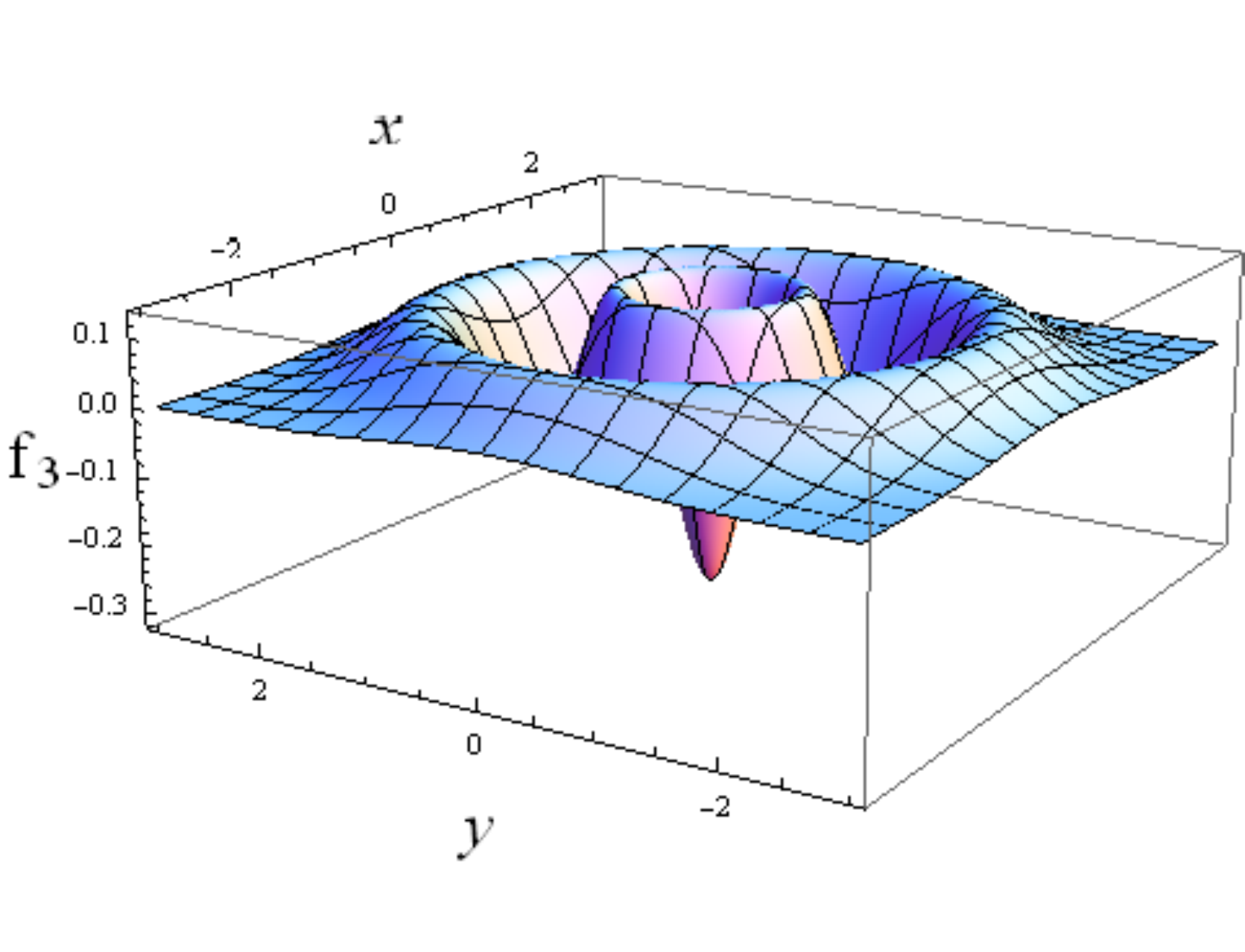}\\[.6ex]
\includegraphics[width=55mm]{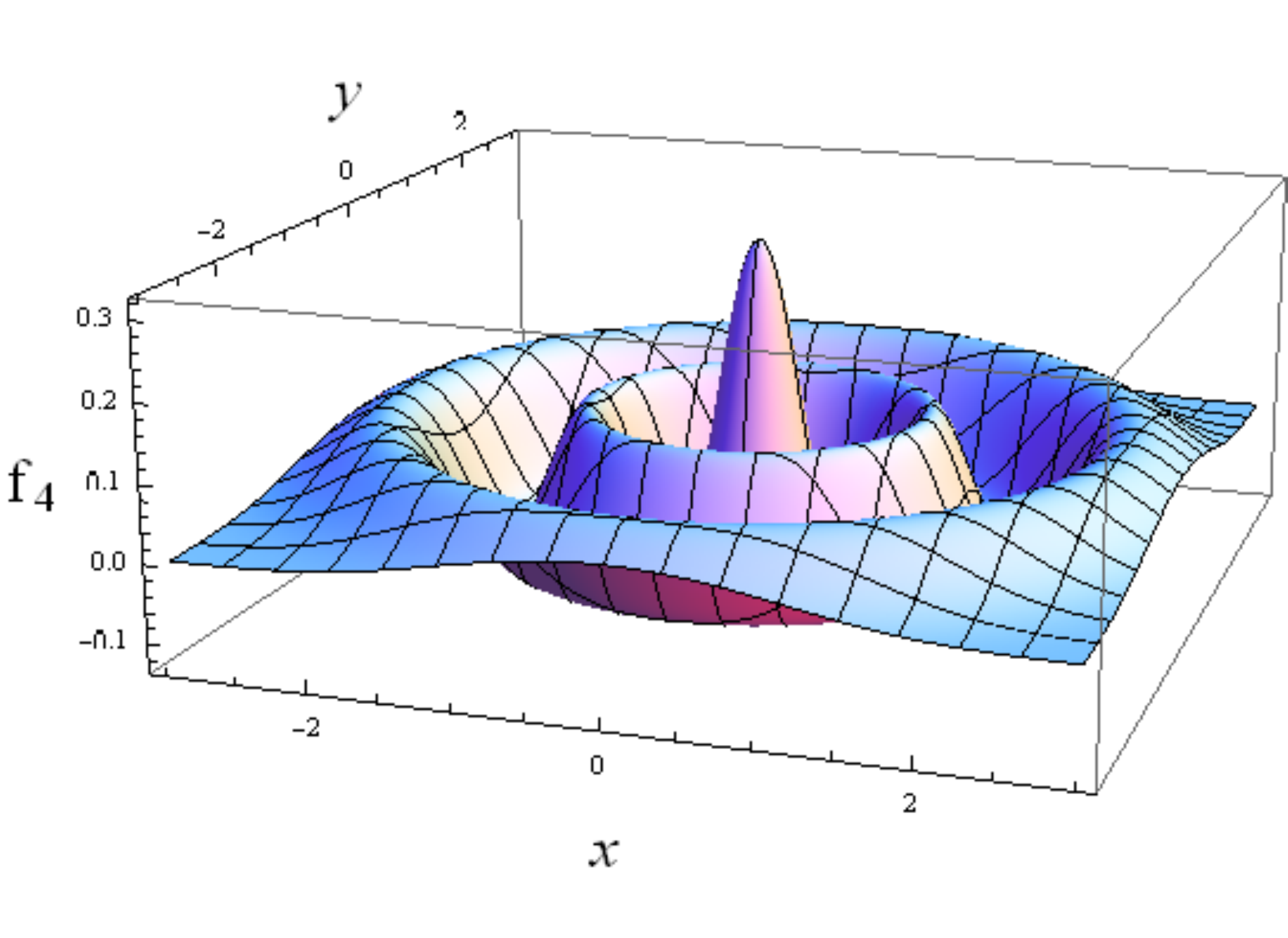}
\end{center}
\caption{Examples minimal information demand/supply surfaces with negative probabilities.}
\label{turbany}
\end{figure}
The preimage of $\mathbb{R}_{-}$ under $f_n$, $n>0$, is the sum of $\lfloor\tfrac{n}{2}\rfloor$ concentric circles for even $n$  and only a single circle  for odd $n$, see Fig.~\ref{turbany}. This is the area where Giffen paradoxes appear\footnote{That is the effect of {\sl turning back} of
the supply and demand curves what often happens for work supplies
and, in general, for the so called  the Giffen goods
\cite{stig,gif}. }. Therefore, minimizing information market strategies besides being gaussian in addition involve bluffing! Violation of laws of supply and demand result from economy (costs reduction)  in information transfer. In this case  $CDF_s$ and $CDF_d$ is not monotonic. This is a consequence of the fact that $f_n$ is negative on some interval (see \cite{Mak}).
One   might by alerted by the appearance of negative probabilities \cite{dirac,feyn,haug} but they are well-defined theoretical concepts, like a negative of  money. 
An interesting model for the study of the negative probability on markets where there are anomalies in the law of supply and demand are presented in \cite{Mak}.

\section{Conclusion}
Why quantum-like description of market make sense? The Fisher information might provide us with the key answer.  For every achievable profit and risk, market adopts the very special strategy $\psi_n$ that discloses only minimal amount of information. The appropriate  total risk =$I_F$ is "quantized" (i.e. takes values from a discrete set) and only for its minimal value  $ \varepsilon\negthinspace=\negthinspace\tfrac{1}{2}$ is compatible with ``market laws'' such as monotonic supply and demand.  The respective time evolution of market can be modelled by mixed strategies \cite{qmg,dyf}. In this case, information loss is measured on the meta-level
by Boltzmann/Shannon entropy \cite{sigmafi}. The resulting quantum-like strategies imply interesting agents' behaviors \cite{PSS}. It is worth emphasizing here that quantum theory imposes desirable properties on player strategies that also characterize liquid goods in the market. Quantum strategies can be identified in a non-destructive way (quantum fingerprinting). 
They also can not be cloned (no-cloning theorem) and its is impossible to delete one of the copies (no-deleting theorem). There are also many other interesting possibilities of using quantum entanglement, cryptography.  For these reasons, it is possible that the quantum market will finally came into existence. 
In this paper, we showed that in such a market only the least risky buying / selling strategies meet the classic laws of supply and demand.

It should be noted that the quantum market model derived from minimal Fisher information can have many inconsistencies with the quantum model of small vibrations of physical systems, giving rise to conclusions on the nature of fundamental phenomena in material systems. These inconsistencies pertain to conclusions referring to market phenomena. For instance, total risk (equivalent of the energy of the material system) does not have to be reduced in market situations involving reasonable players ready to take risks in order to increase transaction profit or in order to deliberately disrupt the market by disinformation caused by using non-monotonic demand/supply strategies (there have been situations where such behaviour was economically justified). Also, the world of trading tactics, whether referring to market agents, humans or automatic systems, should account for factors such as altruism, political market interventions, non-economic benefits, etc. Considering all these factors in a mathematical model is very difficult and continues to be a challenge to researchers.


\end{document}